\documentclass[journal, 10pt]{IEEEtran}
\usepackage{hasan}
\usepackage{authblk}
\usepackage{xcolor}
\usepackage[most]{tcolorbox}
\usepackage{multirow}
\usepackage{tablefootnote}
\usepackage{xspace}
\usepackage{amsmath,amsfonts,amssymb,amsthm, bm}
\newcommand{\dotleq}{\mathrel{\dot{\leq}}}

\newcommand{\rank}[1]{\tr{rank}{\left(#1\right)}}

\newcommand{\zeff}{\mbf z_{\tr{eff}}}
\newcommand{\zvareff}{\sigma_{\tr{eff}}}
\renewcommand{\det}[1]{\tr{det}\left(#1\right)}

\newcommand{\nc}[0]{network coding\xspace}
\newcommand{\Nc}[0]{Network coding\xspace}
\newcommand{\cf}[0]{compute-and-forward\xspace}

\newcommand{\df}[0]{decode-and-forward\xspace}

\newcommand{\af}[0]{amplify-and-forward\xspace}

\newcommand{\marc}{multiple-access relay channels\xspace}

\newcommand{\wre}[1]{\mbf w_{#1}^{\tr{Re}}}
\newcommand{\wim}[1]{\mbf w_{#1}^{\tr{Im}}}
\newcommand{\w}[1]{\mbf w_{#1}}
\newcommand{\ad}[1]{\mbf a_{d_{#1}}}
\newcommand{\ar}[1]{\mbf a_{r_{#1}}}
\newcommand{\whre}[1]{\hat{\mbf w}_{#1}^{\tr{Re}}}
\newcommand{\whim}[1]{\hat{\mbf w}_{#1}^{\tr{Im}}}
\newcommand{\wh}[1]{\hat{\mbf w}_{#1}}
\newcommand{\xre}[1]{\mbf x_{#1}^{\tr{Re}}}
\newcommand{\xim}[1]{\mbf x_{#1}^{\tr{Im}}}
\newcommand{\x}[1]{\mbf x_{#1}}
\newcommand{\ure}[1]{\mbf u_{#1}^{\tr{Re}}}
\newcommand{\uim}[1]{\mbf u_{#1}^{\tr{Im}}}
\newcommand{\uu}[1]{\mbf u_{#1}}
\newcommand{\qre}[1]{q_{#1}^{\tr{Re}}}
\newcommand{\qim}[1]{q_{#1}^{\tr{Im}}}

\newcommand{\vre}[1]{\mbf v_{#1}^{\tr{Re}}}
\newcommand{\vim}[1]{\mbf v_{#1}^{\tr{Im}}}
\newcommand{\vhre}[1]{\hat{\mbf v}_{#1}^{\tr{Re}}}
\newcommand{\vhim}[1]{\hat{\mbf v}_{#1}^{\tr{Im}}}
\newcommand{\uhre}[1]{\hat{\mbf u}_{#1}^{\tr{Re}}}
\newcommand{\uhim}[1]{\hat{\mbf u}_{#1}^{\tr{Im}}}
\newcommand{\uh}[1]{\hat{\mbf u}_{#1}}
\newcommand{\uhd}[1]{\hat{\mbf u}_{d_{#1}}}
\newcommand{\uhr}[1]{\hat{\mbf u}_{r_{#1}}}
\newcommand{\vv}[1]{\mbf v_{#1}}

\newcommand{\cp}{\tr{cp}}

\newcommand{\codlat}{{\Lambda_{\tr{c}}}} 
\newcommand{\shaplat}{\Lambda_{\tr{s}}} 
\newcommand{\Pout}{P_\tr{out}}
\newcommand{\Pdef}{P_\tr{def}}

\newcommand{\Rcp}{R_{\tr{cp}}}
\newcommand{\Rrd}{R_{\tr{rd}}}

\newcommand{\lat}[1]{\Lambda(#1)}

\newcommand{\sucmin}[3]{\lambda_{#1}^{#3}(#2)}

\tikzset{
  curve1/.style={TolDarkBlue, thick, mark=*, mark size=2.4pt, mark options=solid},
  curve2/.style={TolLightBrown, thick, mark=square*, mark size=2.2pt, mark options=solid},
  curve3/.style={TolLightGreen, thick, mark=triangle*, mark size=3.0pt, mark options=solid},
  curve4/.style={TolDarkBrown, thick, mark=diamond*, mark size=2.9pt, mark options=solid},
  curve5/.style={TolLightPink, thick, mark=ball, mark size=2.4pt, mark options=solid},
  curve6/.style={TolIndigo, thick, mark=oplus, mark size=2.4pt, mark options=solid},
  curve7/.style={TolGrey, thick, mark=pentagon, mark size=2.4pt, mark options=solid},
}

\makeatletter
\newcommand{\removelatexerror}{\let\@latex@error\@gobble}
\makeatother
\IEEEoverridecommandlockouts

\begin{document}
\title{Opportunistic Cooperation Strategies for Multiple Access Relay Channels with Compute-and-Forward}
\author{Mohammad~Nur~Hasan,~\IEEEmembership{Student Member,~IEEE,}
  Brian~M.~Kurkoski,~\IEEEmembership{Member,~IEEE}
  \thanks{Mohammad Nur Hasan and Brian M. Kurkoski are with Japan Advanced Institute of Science and Technology (JAIST).}
  \thanks{Our work related closely to this paper were presented in the 2017 IEEE International
      Conference on Communications (ICC) \cite{HasanICC17} and the 2018 International Symposium on Information Theory and Its Applications (ISITA) \cite{HasanISITA18}.}

}

\maketitle

\begin{abstract}
  This paper studies the application of compute-and-forward to multiple access
  relay channels (MARC). Despite its promising advantage of improving network
  throughput, it is not straightforward to apply compute-and-forward to the
  MARC. This paper proposes two efficient cooperation strategies for the MARC
  with compute-and-forward. Both proposed strategies are opportunistic in the
  sense that the cooperation between the relay and the destination are performed
  only when it is needed for the sake of high transmission efficiency. In the
  first strategy, the relay helps the destination by sending its local optimal
  integer coefficient vector without taking into account that the resulting integer
  coefficient matrix at the destination is full rank. In contrast, in the second
  strategy, the relay forwards its ``best'' coefficient vector that always ensures a full rank coefficient matrix at the destination. Both of the proposed strategies achieve twice the network throughput of the existing strategies at high signal-to-noise power ratio (SNR). They also have lower outage probability, independent of relay placement. Furthermore, the first strategy nearly achieves diversity gain of order two, and the second one achieves exactly the diversity gain of order two, which cannot be achieved by the existing strategies.
\end{abstract}


\section{Introduction}
\label{sec:introduction}

\Nc \cite{AhlswedeCLY00,ZhangPLNC,SykoraBurr18,WilsonNPS10} has become an
important networking strategy to improve the spectral efficiency of wireless
communication networks. In contrast to simple forwarding, \nc allows
intermediate nodes to ``combine'' the received  messages before forwarding them to
following nodes, to reduce the required number of transmissions. On the other hand,
cooperative communication is an effective method to enlarge network coverage,
increase transmission robustness, and improve power efficiency by exploiting
spatial diversity without additional antennas
\cite{CoverG79,LanemanTW04,Gamal12}. However, the gains achieved by cooperative
communications in practice come with a loss of spectral efficiency due to
half-duplex operation \cite{LanemanTW04}. Thus, it is beneficial to apply \nc to
cooperative networks to achieve reliable communications with high spectral
efficiency. 

In this paper, we design network coding schemes for the multiple access relay
channel (MARC), which is an important class of wireless cooperative networks. 
In the MARC, multiple sources want to deliver messages to one common destination
with the assistance of one relay node
\cite{KramerW00,HauslD06,Wodldegebreal07,Wei15,HeTad18}.
The applications of
such networks include sensor and ad-hoc networks and uplink for cellular
networks with an intermediate node as a relay. The MARC also has found a use case in
the \textit{LTE Advanced} mobile communication standard \cite{Insausti2019,
  lteAdvancedMarc}.
It has been shown that \nc significantly improves the spectral efficiency of the MARC. For example, in the
conventional two-source MARC, four orthogonal transmissions are required,
where the sources transmit their messages in turn and the relay forwards the
messages one by one to the destination. With \nc, the number of transmissions is
reduced to only three; the first and the second transmissions are used by the
sources to transmit messages in turn, while the third transmission is used by
the relay to forward the \textit{network coded} version of the transmitted
messages to the destination \cite{HauslD06}. 

Recently, Nazer and Gastpar proposed a new network coding and relaying scheme,
known as \cf \cite{NazerG11}. It views interference as an advantage to exploit rather than
a problem to avoid, and allows sources in a relay network to
simultaneously transmit their messages via a non-orthogonal channel. Each relay
directly \textit{computes} an integer linear combination of the transmitted
messages from the received superimposed signals without decoding each
transmitted message separately, and then forwards it to the destination. Given a
sufficient number of linear combinations, the destination can recover the
transmitted messages so long as the coefficient matrix, that is the matrix
composed of the coefficients of the linear combinations, is full rank. 

Allowing sources to transmit their messages via one non-orthogonal channel is
an appealing advantage of \cf. It is easy to see its potential for improving the
spectral efficiency of the MARC. When \cf is applied to the MARC, the required number of
transmissions can be reduced to only two; the first is used by the sources
to broadcast their messages to the relay and destination at once and the second
is used by the relay to forward its computed linear combination to the
destination. Note that this advantage is applicable to any MARC with any number of
sources, not limited to the MARC with two sources. Despite its promising
advantage, however, it is not straightforward to efficiently apply \cf to MARC
as the destination requires the resulting coefficient matrix to be full rank.

It is possible to naively apply the original \cf \cite{NazerG11} to the MARC by
allowing the destination and the relay to compute their local optimal integer
linear combinations independently. However, it may result in a rank deficient
coefficient matrix which causes a decoding failure. Therefore,  
cooperation between the destination and the relay in computing their integer
linear combinations is necessary. In \cite{SoussiZV14}, Soussi \textit{et al.}
made an attempt to solve this issue. They proposed a global optimization
technique such that the destination and the relay select the global optimal
linearly independent combinations with respect to achievable rate. They showed that with this
strategy, \cf achieves higher achievable symmetric-rate compared to other
relaying strategies such as \df and \af. The achievable rate improvement in their
work, however, relies on the assumption that all channel state information (CSI) are known to all
nodes. Insausti \textit{et al.} \cite{Insausti2019} proposed another strategy for applying \cf to the
MARC. Distinct from \cite{SoussiZV14}, the relay is allowed to select its local
best linear combination; based on this the destination adjusts its linear
combination ensuring a full rank coefficient matrix. This strategy is more efficient than the one proposed in
\cite{SoussiZV14}, and also has a higher achievable
rate. Both of the aforementioned works focused on
the achievable rate performance without investigating outage probability performance. Because
one of the main objectives of wireless cooperative networks is to increase
transmission reliability, it is important to make sure that the outage probability
performance is also improved when \cf is employed.  In this work we design efficient
cooperation strategies for the MARC employing \cf that improve outage probability performance as well. 

The main contributions of this paper are the two efficient cooperation strategies proposed for applying \cf to the MARC. While improving the outage probability performance, the proposed strategies utilize transmission channels efficiently and allow the relay to help the destination only when necessary. In the first strategy, the relay helps the destination by forwarding its local best linear combination without taking into account the possibility that the resulting integer coefficient at the destination is not full rank. Contrarily, in the second strategy, the relay forwards its best linear combination that always ensures a full rank integer coefficient matrix at the destination. Semi-theoretical and numerical analyses are provided to show the performance of the proposed strategies in terms of outage probability, diversity gain, and network throughput. It is shown that independent of relay placement, the proposed strategies always have lower outage probability compared to the Soussi \cite{SoussiZV14} and Insausti \cite{Insausti2019} strategies. It is also revealed that the first strategy nearly achieves diversity gain of order two, which is a significant improvement over the Soussi strategy \cite{SoussiZV14}. On the other hand, the second strategy exactly achieves the diversity gain of order two, which is the full diversity gain of the MARC.\footnote{In this paper, because all sources transmit independent messages, the full diversity gain of the MARC is of order two.} Moreover, the results show that both of the proposed strategies yields twice the throughput of the existing strategies \cite{SoussiZV14} and \cite{ Insausti2019} at high signal-to-noise power ratio (SNR).

\textit{Notation:} $\mbb{R}$, $\mbb{C}$, and $\mbb{Z}$ denote the real, complex,
and integer numbers, respectively. $\mbb{F}_p$ represents the finite field of
size $p$, where $p$ is a prime number. The Gaussian integers are denoted by
$\mbb{Z}[i]$. Boldface lowercase letters denote vectors, e.g., $\mathbf{a} \in
\mathbb{Z}[i]^m$, while boldface uppercase letters denote matrices, e.g.,
$\mathbf{A} \in \mathbb{Z}[i]^{m \times m}$. The identity matrix of size $m$ is
denoted as $\mbf{I}_m$. For a vector $\mbf{a}$, we use $a_i$ to denote the
element with index $i$. For a matrix $\mbf{A}$, we denote the element at row $i$
and column $j$ as $a_{ij}$. The rank and deteminant of a matrix $\mbf A$ are denoted as $\rank{\mbf A}$ and $\det{\mbf A}$, respectively. The Hermitian transpose and the regular transpose
are expressed by superscripts ``$H$'' and ``$T$'', e.g.,  $\mbf{A}^H$ and
$\mbf{A}^T$, respectively. The
$\log$ operation is with respect to base $2$ and $\log^{+}(x) \triangleq \max
(\log(x), 0)$. The real and imaginary components of a complex number are denoted
using $\Re(\cdot)$ and $\Im(\cdot)$, respectively.  
\section{Multiple Access Relay Channel Model}
\label{sec:mult-access-relay}

\begin{figure}[ht]
 \centering
  \begin{tikzpicture}[mynode/.style={draw, circle, inner sep=8pt},
    myarrow/.style={->, >=stealth},
    scale=1.0,
    ]
    \node[mynode] (R) at (0,0) {};
    \node[label=below:Relay] at (R) {$ r$};
    \node[mynode] (D) at (3,0) {};
    \node[label=below:Destination] at (D) {$ d$};
    \node[mynode] (S1) at (-3,2) {};
    \node at (S1) {$ s_1$};
    \node[mynode] (S2) at (-3,1) {};
    \node at (S2) {$ s_2$};
    \fill (-3,0) circle (1pt) ++(0,-0.5) circle(1pt) ++(0,-0.5) circle(1pt);
    \node[mynode] (SK) at (-3,-2) {};
    \node[label=below:Sources] at (SK) {$ s_M$};
    \foreach \x/\y in {S1/R, S1/D, S2/R, S2/D, SK/R, SK/D}
    \draw[myarrow, TolIndigo] (\x) -- (\y);
    \draw[myarrow, TolDarkBrown, densely dashed] (R) -- (D);
    \draw[<->, densely dashdotted] (-3, 2.5) -- node[midway, below]{$\delta_{sd} = 1$} (3, 2.5);
    \draw[<->, densely dashdotted] (-3, 2.7) -- node[midway, above]{$\delta_{sr}$} (-0.05, 2.7);
    \draw[<->, densely dashdotted] (0.05, 2.7) -- node[midway, above]{$\delta_{rd}$} (3, 2.7);
    \coordinate [label=right:{\color{TolIndigo}{trans. round 1}}] (label1) at (2,-1.5);
    \coordinate [label=right:{\color{TolDarkBrown}{trans. round 2}}] (label2) at (2,-2);
    \draw[myarrow, TolIndigo] (1, -1.5) -- (label1);
    \draw[myarrow, TolDarkBrown, densely dashed] (1, -2) -- (label2);
  \end{tikzpicture}
  \caption{$M$-users MARC model.}
  \label{fig:system-model}
\end{figure}
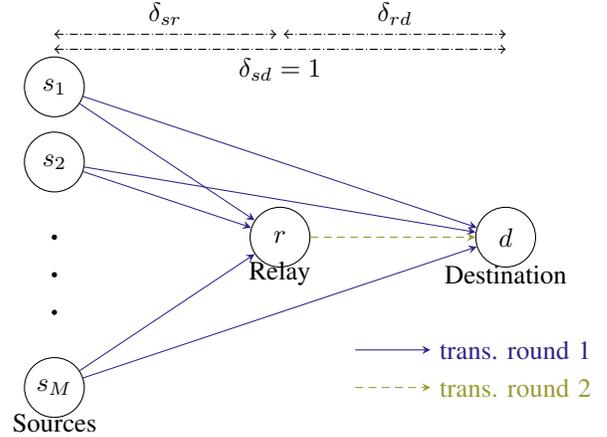

We begin by describing the system model of the MARC
considered in this paper. As illustrated in Fig.~\ref{fig:system-model}, our
model consists of $M$ sources denoted as $s_m$, $m = 1, \dots, M$, one
destination $d$ and one relay node $r$. The sources want to transmit
information messages to the destination. All wireless links are assumed to 
be Rayleigh block fading channels where the fading coefficients remain constant
within a block of symbols, but change independently from one block to the other
according to a circularly symmetric complex Gaussian distribution with zero mean
and unit variance. 

As shown in Fig.~\ref{fig:system-model}, there are three types of directed transmission
links: sources-destination, sources-relay, and relay-destination links. All
transmitters (sources and relay) have the same average transmit power
$P$. For $i \in \{s_1,\dots,s_M,r\}$, $j \in \{r,d\}$, and $i \neq j$, let
$\gamma_{ij}$, $\delta_{ij}$, $g_{ij}$, and $h_{ij}$ be the average
SNR, distance, geometric gain, and channel gain of
wireless link from node $i$ to node $j$, respectively. The geometric
gain captures the effect of path loss which is a function of distance, i.e.,
$g_{ij} = \delta_{ij}^{-\kappa}$, where $\kappa$ is the path loss exponent
\cite{HolmaWCDMA,Youssef11}. The channel gain $h_{ij}$  and noise at every node
is randomly distributed over $\mcal{CN} (0,1)$. Thus, the average SNR can be
defined as $\gamma_{ij} \triangleq Pg_{ij}$. For simplicity, we assume that  all
sources have the same distance to the destination and also have the same
distance to the relay, To be more precise, let $\delta_{sd}$ and $\delta_{sr}$
be positive real scalars. We assume that for all $m \in \{1, \dots, M\}$,
$\delta_{s_md} = \delta_{sd}$ and $\delta_{s_mr} = \delta_{sr}$.  This
assumption implies that the sources have one common geometric gain to the
destination and another common geometric gain to the relay, i.e.,  $\forall m$,
$g_{s_md} = g_{sd}$ and $g_{s_mr} = g_{sr}$ for positive scalars $g_{sd}$ and $g_{sr}$. Accordingly, the average SNR of
each source to the destination can be defined as $\gamma_{s_md} = \gamma_{sd} = Pg_{sd}$, and
to the relay as $ \gamma_{s_mr} = \gamma_{sr} = Pg_{sr}$.

\section{MARC with Compute-and-Forward}
\label{sec:marc-with-cf}
In this section we provide an exposition of compute-and-forward and how it is adopted to the MARC. See \cite{NazerG11} for in-depth discussion of compute-and-forward.
\subsection{Encoding Scheme}
\label{sec:marc-encoding}

For simplicity, we consider a symmetric MARC where all sources transmit with the
same rate $R$. For a positive integer $n$, consider nested lattices $\shaplat
\subseteq \codlat \subset \mbb R^n$ and let $\mcal C
\triangleq \mcal C(\codlat / \shaplat)$ be a nested lattice code with rate
$R/2$.\footnote{In-depth discussion about nested lattice codes can be found in \cite{ConwayS99,ZamirSE02,Kurkoski16}} $\codlat$ represents a fine lattice used for
coding and $\shaplat$ represents a coarse lattice used for shaping to ensure that the
average power constraint is satisfied. The second moment of $\shaplat$ is assumed to be 
$P/2$. For a prime number $p$, let $\mcal E$ be a bijective mapping from $\mbb F_p^{nR/2}$ to
$\mcal C$, i.e.,
\begin{align}
  \mcal E : \mbb F_p^{nR/2} \rightarrow \mcal C.
\end{align}
The bijective mapping $\mcal E$ will be employed by the sources and the relay for encoding their messages to lattice codewords. 

The encoding is performed as follows. Each source $s_m$ randomly generates two information vectors $\wre m, \wim m \in
\mbb F_p^{nR/2}$. Together, these information vectors form $ \w m = (\wre m, \wim m) \in
\mbb F_p^{nR}$, which are then encoded to a complex-valued vector in the following way.
$\wre m$ and $\wim m$ are respectively mapped  to $\xre m \in \mcal C$ and
$\xim m \in \mcal C$ using $\mcal E$, i.e.,  
\begin{align}
  \label{eq:marc-encoding}
  \xre m = \mcal E (\wre m),\\
  \xim m = \mcal E (\wim m).
\end{align}
Subsequently, these real-valued vectors are used to form complex-valued vectors $\x m = \xre m + i \xim m \in \mbb C^n $ which
are broadcast to the destination and the relay. We assume
that a dithering technique \cite{ErezZ04Awgn} is employed. However, we omit the description for ease exposition. Dithering is important for ensuring the resulting
effective noise is independent of the underlying lattice codewords. Furthermore,
it ensures that each $\x m$ satisfies the average power
constraint $\mbb E \{ \norm{\x m }^2 \} \leq nP$. 

\subsection{Transmission Rounds}
\label{sec:marc-transmission}
The end-to-end information transmission is divided into two rounds. In the first
round, the sources simultaneously broadcast $\x m$ to the
destination and the relay. For $j \in \{r,d\}$, let $\mbf z_j$ be a noise vector
at node $j$, and recall that $h_{ij}$ denotes the channel coefficient from
node $i$ to $j$, $i \in \{s_1,\dots,s_M\}$. The destination and the relay
respectively receive noisy superposition signals
\begin{align}
  \mbf y_d^{(1)} &= \sum_{m=1}^M \sqrt{g_{sd}} h_{s_md}\mbf x_m + \mbf z_d^{(1)}, \label{eq:marc-first-round-destination}\\
  \mbf y_r^{(1)} &= \sum_{m=1}^M \sqrt{g_{sr}} h_{s_mr}\mbf x_m + \mbf z_r^{(1)}.\label{eq:marc-first-round-relay}
\end{align}

At the end of the first round, the relay does not attempt to decode $\w
1,\dots,\w M$ separately as usually done in conventional MARC schemes. Rather, it adopts
the \cf technique to directly decode a linear combination of $\w 1, \dots, \w M$.
Let $\mbf u_r \triangleq f_r(\w 1, \dots, \w M)$ be the desired linear
combination. In the second round,  the relay encodes $\mbf u_r$ to a
complex-valued vector $\x r \in \mbb C^n$ using the same encoding scheme described in
Subsection~\ref{sec:marc-encoding} and forwards it to the destination. We
shall note that to increase the transmission efficiency, the second round is
only used when certain conditions are met, which will be discussed further in
Section~\ref{sec:prop-coop-stat}. The received signal at
the destination is given by 
\begin{align}
  \label{eq:receive-signal-round2}
  \mbf y_d^{(2)} &= \sqrt{g_{rd}} h_{rd}\mbf x_r + \mbf z_d^{(2)}.
\end{align}

\subsection{Computing Linear Combinations}
\label{sec:marc-comp-lin-comb}
As mentioned above, by the end of the first transmission round, the relay
decodes a combination of the transmitted information messages. In fact, it is not only the relay.
Because at least $M$ linear combinations are required to recover all the transmitted
information messages, the destination also needs to decode at least $M-1$ linear
combinations. For simplicity, let us focus on how a receiver, which may represent
either the destination or the relay, decodes some linear combinations.

Before going further, let us rewrite the received signals \eqref{eq:marc-first-round-destination} or
\eqref{eq:marc-first-round-relay} in a simpler form, omitting the notations for
the relay or the destination. Let $\mbf h = [h_1, \dots, h_M] \in \mbb C^M$ be
the channel coefficient vector and $g$ be the geometric gain from the sources to the
receiver. The received signal is rewritten as
\begin{align}
  \label{eq:marc-received-signal}
  \mbf y &= \sum_{m=1}^M \sqrt{g} h_{m}\mbf x_m + \mbf z. 
\end{align}
Assume that the receiver expects to decode $L \leq M$ linear combinations $\uu 1,
\dots, \uu L \in \mbb F_p^{nR}$. For $l \in \{1, \dots, L\}$, the
receiver selects coefficients $\qre {lm}, \qim {lm} \in \mbb F_p$ and attempts
to decode two equations
\begin{align}
  \ure {l}  &= \bigoplus_{m=1}^M  \qre {lm} \wre m \oplus (- \qim {lm}) \wim {m} ,\label{eq:ure-marc}\\
  \uim {l}  &= \bigoplus_{m=1}^M  \qim {lm} \wre m \oplus (\qre {lm}) \wim {m} ,\label{eq:uim-marc}
\end{align}
where $(-q_{m})$ denotes the additive inverse of $q_{m}$. $\bigoplus$ and $\oplus$ denote the summation and addition in $\mbb F_p$, respectively. The linear combinations $\uu l$ are then obtained as $\uu l \triangleq (\ure l,
\uim l)$.

Although the desired linear combinations are evaluated over the finite field
$\mbb F_p$, the channels operate over the complex field $\mbb C$. This issue
can be overcome by exploiting the real-valued decomposition of a complex-valued number.
To be precise, the receiver selects an integer linear coefficient $\mbf
a_{l} = [a_{l1}, \dots, a_{lM}] \in \mbb Z[i]^{M}$ and decodes the corresponding lattice equation
\begin{align}
  \label{eq:marc-lat-eq}
  \vv {l} = \sum_{m=1}^M a_{lm} \x m \bmod \shaplat.
\end{align}
Now, $\vv {l}$ can be written as $\vv {l} = \vre {l} + i
\vim {l}$, where  
\begin{align}
  \vre {l} &\triangleq \Re (\vv {l}) = \left[ \sum_{m=1}^M \Re (a_{lm}) \xre m - \Im (a_{lm}) \xim m \right] \bmod \shaplat \label{eq:vre-marc-lat-eq}\\
  \vim {l} &\triangleq \Im (\vv {l}) = \left[ \sum_{m=1}^M \Im (a_{lm}) \xre m + \Re (a_{lm}) \xim m \right] \bmod \shaplat. \label{eq:vim-marc-lat-eq}
\end{align}
Once the destination obtains $\vv {l}$, it can recover $\ure {l}$ and
$\uim {l}$ using the inverse of $\mcal E$ as follows
\begin{align}
  \ure {l} \triangleq \mcal E^{-1} (\vre {l}) &= \bigoplus_{m=1}^M \qre {lm} \wre m \oplus (- \qim {lm}) \wim {m} ,\label{eq:ure-marc-mapped}\\
  \uim {l} \triangleq \mcal E^{-1} (\vim {l}) &= \bigoplus_{m=1}^M \qim {lm} \wre m \oplus (\qre {lm}) \wim {m} .\label{eq:uim-marc-mapped}
\end{align}
Given the choice of $\mbf a_{l}$, $\qre {lm}$ and $\qim {lm}$
are equivalent to $\qre {lm} = \Re(a_{lm}) \bmod p$ and $\qim {lm} = \Im(a_{lm})
\bmod p$. This implies that the selection of integer coefficients in the Gaussian
integer domain corresponds to the selection of coefficients in the finite field
domain.

Now, in order to obtain the lattice equation $\mbf v_l$, the receiver first scales $\mbf y$ with a scaling factor $\alpha_l \in \mbb C$ and
computes
\begin{align}
  \label{eq:marc-rec-sig-scaled}
  \tilde{\mbf y}_l &= [\alpha_l \mbf y ] \bmod \shaplat \\
                   &= \left[ \sum_{m=1}^M \alpha_l \sqrt{g} h_m \x m + \alpha_l \mbf z \right] \bmod \shaplat \\
                   &= \Big[ \sum_{m=1}^M a_{lm} \x m + \sum_{m=1}^M \left(  \alpha_l \sqrt{g} h_m \x m - a_{lm} \x m \right) \nonumber \\
                   & \quad + \alpha_l \mbf z \Big] \bmod \shaplat \nonumber \\
                   &= \left[ \vv l + \zeff (\alpha_l, \mbf a_l, \mbf h, g)  \right] \bmod \shaplat,
\end{align}
where
\begin{align}
  \label{eq:marc-zeff}
  \zeff (\alpha_l, \mbf a_l, \mbf h, g)= \sum_{m=1}^M \left(  \alpha_l \sqrt{g} h_m \x m - a_{lm} \x m \right) + \alpha_l \mbf z
\end{align}
is the effective noise. Subsequently, it produces estimates of $\vre l$ and $\vim
l$ by quantizing the real and imaginary components of $\tilde{\mbf y}_l$ with
respect to $\codlat$ and performs the modulo operation with respect to
$\shaplat$, i.e.,
\begin{align}
  \vhre{l} &= Q_{\codlat} \left( \Re(\tilde{\mbf y}_l) \right) \bmod \shaplat \label{eq:marc-quant-mod-re} \\
  \vhim{l} &= Q_{\codlat} \left( \Im(\tilde{\mbf y}_l) \right) \bmod \shaplat.\label{eq:marc-quant-mod-im}
\end{align}
Finally, the estimates of $\ure l$ and $\uim l$ are obtained using the
inverse of $\mcal E$,
\begin{align}
  \uhre l &= \mcal E^{-1} \left( \vhre l \right), \label{eq:marc-inv-enc-re} \\
  \uhim l &= \mcal E^{-1} \left( \vhim l \right).\label{eq:marc-inv-enc-im}
\end{align}
The estimate of $\uu l$ is then recovered as $\hat{\mbf u}_l =
\left( \uhre l, \uhim l \right)$.

In order for the receiver to be able to decode $\uu l$ with low error
probability, the scaling factor $\alpha_l$ has to be chosen such that the
variance of the effective noise $\zeff(\alpha_l, \mbf a_l, \mbf h, g)$ is
minimized. Let $\zvareff^2(\alpha_l, \mbf a_l, \mbf h, g)$ be the variance of $\zeff(\alpha_l, \mbf a_l, \mbf h, g)$ defined as
\begin{align}
  \label{eq:marc-zeff-var}
  \zvareff^2(\alpha_l, \mbf a_l, \mbf h, g) &\triangleq \frac 1 n \mbb E \left\{ \norm{\zeff(\alpha_l, \mbf a_l, \mbf h, g)}^2 \right\} \\
                                         &= \norm{\alpha_l \sqrt{g} \mbf h - \mbf a_l}^2 P + \abs{\alpha_l}^2.
\end{align}
One can show that the optimal value for $\alpha_l$ is given by
\begin{align}
  \label{eq:marc-opt-scaling}
  \alpha_l^{\tr{opt}} &\triangleq \argmin_{\alpha_l}  \zvareff^2(\alpha_l, \mbf a_l, \mbf h, g) \\
                      &= \frac {P \sqrt{g} \mbf h^H \mbf a_l} { 1 + P g \norm{\mbf h}^2}.
\end{align}

As a summary, the receiver decodes linear combination $\uu l$ in the following
way. It first selects an integer coefficient vector $\mbf a_l \in \mbb Z[i]^M$,
then computes $\alpha_l^{\tr{opt}}$ and uses it as the scaling
factor $\alpha_l$. Next, it scales the received signal and performs the modulo
operation with respect to $\shaplat$. Finally, the desired linear combination is
obtained by performing operations described in
\eqref{eq:marc-quant-mod-re}, \eqref{eq:marc-quant-mod-im}, \eqref{eq:marc-inv-enc-re},
and \eqref{eq:marc-inv-enc-im} sequentially.

Next, we discuss how to choose the integer coefficient vector $\mbf a_l$. In principle, it is possible for the receiver to choose any integer coefficient
vector. However, the selection of the coefficient vector has a significant impact on
the achievable computation rate and consequently on the outage probability.
Therefore, $\mbf a_l$ has to be chosen carefully.
The achievable computation rate of \cf in complex-valued channels is given in the following theorem. 

\begin{mytheorem}[Computation Rate \cite{NazerG11}]
  \label{eq:comp-rate-complex}
  Consider a complex-valued Gaussian network with $M$ transmitters that simultaneously transmit
  their messages with average power constraint $P$ to a receiver. Let $\mbf h =
  [h_1,\dots,h_M]^T \in \mbb C^M$ be the channel coefficients and $g$ be the
  geometric gain from the transmitters to the receiver. Given a coefficient vector $\mbf
  a = [a_1, \dots , a_M] \in \mbb Z[i]$, the receiver can decode the corresponding 
  linear combination of transmitted messages  with low error probability so long as
  the message rate is less than the \textit{computation rate} given by
  \begin{align}
    \label{eq:comp-rate-complex}
    R_{\cp}(\mbf a, \mbf h, g)=  \log^+ \left( \left( \norm{\mbf a}^2 - \frac {P g \abs{ \mbf h^H \mbf a }^2}{1 + P g \norm{\mbf h}^2} \right)^{-1} \right). 
  \end{align}
\end{mytheorem}

From the above theorem, it is clear that the selected integer coefficient $\mbf a$ is a component that determines the computation rate. Therefore, it should be chosen such that the computation rate is maximized. 

For $l \in
\{1,\dots, L\}$, let $\mbf a_l$ be the integer coefficient vector of the
desired linear combination $\uu l$. Let ${\mbf A} = [
\mbf a_1, \dots , \mbf a_L]^T$ and $\mbf a_1, \dots, \mbf a_L$ should be chosen to be linearly
independent, and thus, $\rank{{ \mbf A }} = L$. The receiver selects $\mbf A$ such that
\begin{align}
  \label{eq:marc-A-opt}
  { \mbf A } = \argmax_{\substack{\tilde{ \mbf A } = [\tilde{\mbf a_1}, \dots , \tilde{\mbf a_L}] \in \mbb Z[i]^{L \times M},\\ \rank{\tilde{ \mbf A }} = L}} \; \min_{l=1,\dots,L} R_{\cp}(\tilde{\mbf a_l}, \mbf h, g).
\end{align}

It can be shown that the computation rate $R_{\cp}(\mbf a_l, \mbf
h, g)$ can be written as \cite{FengSK13}
\begin{align}
  \label{eq:marc-r-comp-simplified}
  R_{\cp}(\mbf a_l, \mbf h, g) = \log^+ \left( \frac 1 {\mbf a_l^H \mbf M \mbf a_l } \right),
\end{align}
where
\begin{align}
  \label{eq:marc-M-mat}
  \mbf M = \mbf I - \frac {Pg}{1 + Pg \norm{\mbf h}^2} \mbf {hh}^H.
\end{align}
One can observe that $\mbf M$ is a positive definite matrix, and thus, using
Cholesky factorization, it can be decomposed into
\begin{align}
  \label{eq:marc-chol-decomp}
  \mbf M = \mbf B^H \mbf B,
\end{align}
where $\mbf B$ is an upper triangular matrix.

The problem defined in \eqref{eq:marc-A-opt} now can be transformed into
\begin{align}
  { \mbf A } &= \argmax_{\substack{\tilde{ \mbf A } = [\tilde{\mbf a_1}, \dots , \tilde{\mbf a_L}] \in \mbb Z[i]^{L \times M},\\ \rank{\tilde{ \mbf A }} = L}} \; \min_{l=1,\dots,L} \log^+ \left( \frac 1 {\tilde{\mbf a_l}^H \mbf M \tilde{\mbf a_l} } \right) \\
             &= \argmin_{\substack{\tilde{ \mbf A } = [\tilde{\mbf a_1}, \dots , \tilde{\mbf a_L}] \in \mbb Z[i]^{L \times M},\\ \rank{\tilde{ \mbf A }} = L}} \; \max_{l=1,\dots,L} \tilde{\mbf a_l}^H \mbf M \tilde{\mbf a_l}  \\
             &= \argmin_{\substack{\tilde{ \mbf A } = [\tilde{\mbf a_1}, \dots , \tilde{\mbf a_L}] \in \mbb Z[i]^{L \times M},\\ \rank{\tilde{ \mbf A }} = L}} \; \max_{l=1,\dots,L} \norm{\mbf B \tilde{\mbf a_l}}^2. \label{eq:marc-A-opt-suc-min} 
\end{align}

\begin{mydefinition}[Successive Minima]\label{def:successive-minima}
  For an $n$-dimensional lattice $\lat{ \mbf G}$, the $l$-th successive minimum, $1 \leq l \leq n$, is defined as
  \begin{align}
    \sucmin{l}{\mbf G}{} \triangleq \min_{\mbf v_1,...,\mbf v_l \in \lat{\mbf G}} \max \{\norm{\mbf v_1}, ..., \norm{\mbf v_l}\}, \label{eq:successive-minima}
  \end{align}
  where the minimum is taken over all sets of $l$ linearly independent vectors in $\lat{\mbf G}$. In other words, $\sucmin{l}{\mbf G}{}$ is the smallest real number $r$ such that there exist $l$ linearly independent vectors $\mbf v_1, ..., \mbf v_l \in \lat{\mbf G}$ with $\norm{\mbf v_1}, ..., \norm{\mbf v_l} \leq r$. 
\end{mydefinition}

From \eqref{eq:marc-A-opt-suc-min} and Definition~\ref{def:successive-minima}, it can be said that finding $L$ 
``best'' integer coefficient vectors $\mbf a_1, \dots, \mbf a_L$ with respect to
the computation rate is equivalent to finding integer vectors providing $L$
successive minima of the lattice with a generator matrix $\mbf B$. Thus, to find ${\mbf A}$, we can employ
existing algorithms for finding the successive minima of a lattice such as
the Fincke-Pohst algorithm \cite{FinckeP85}, the Schnorr-Euchner algorithm
\cite{SchnorrE94}, the LLL algorithm \cite{Lenstra82LLL}, and their variations
\cite{ViterboB99SD,DingSMPIF15,WenSMP19,Agrell02,Liu16,SakzadHV13}. 

\subsection{Recovering Information Messages}
\label{sec:marc-recover-trans-msg}

We have discussed that upon receiving the noisy superposition signal in
\eqref{eq:marc-first-round-destination} and \eqref{eq:marc-first-round-relay},
the destination and the relay compute linear combinations of the transmitted messages.
However, the ultimate goal of the destination is to recover the transmitted
information messages $\w 1, \dots, \w M$. To this end, the destination requires
at least $M$ linear combinations. Let us assume that the destination possesses
linear combinations $\uh 1, \dots, \uh M$. These linear combinations may be obtained
either directly by the destination itself, or with the help of the relay. How these linear combinations
are obtained will be addressed in the next section.

Let $\mbf a_1, \dots, \mbf a_M \in \mbb Z[i]^M$ be the integer coefficient
vectors corresponding to $\uh 1, \dots, \uh M$ and let ${\mbf A} = [\mbf a_1,
\dots, \mbf a_M]^T$; we refer to $\mbf A$ as  \textit{integer coefficient
  matrix} or just \textit{coefficient matrix}. The corresponding coefficient matrix in $\mbb F_p$ can be written as
\begin{align}
  \label{eq:marc-Q-mat}
  \mbf Q =
  \begin{bmatrix}
    \Re ({\mbf A}) & -\Im ({\mbf A})\\
    \Im ({\mbf A}) & \Re ({\mbf A})
  \end{bmatrix} \bmod p,
\end{align}
where the modulo operation is element-wise. For all $m \in \{1,\dots,M\}$, let
$\wh m = (\whre m, \whim m)$ be the estimate of $\w m$. The destination decodes
the transmitted information messages by solving the following linear equation
\begin{align}
  \label{eq:final-decoding}
  \begin{bmatrix}
    \uhre 1 \\
    \vdots \\
    \uhre M \\
    \uhim 1 \\
    \vdots \\
    \uhim M
  \end{bmatrix}
  = \mbf Q
  \begin{bmatrix}
    \whre 1 \\
    \vdots \\
    \whre M \\
    \whim 1 \\
    \vdots \\
    \whim M
  \end{bmatrix}
\end{align}
where all operations are performed over finite field $\mbb F_p$. It should be
noted that the destination can solve the above linear equation system if and
only if the matrix $\mbf Q$ is full-rank. Therefore, we should take into account
the probability that the coefficient matrix is not full rank when designing
the MARC with \cf. Nevertheless, it has been shown in \cite[Sec. VI]{NazerG11} that by
taking the blocklength $n$ and field size $p$ to be large enough, rather than
checking the rank of $\mbf Q$ over $\mbb F_q$, it is sufficient to check whether $\mbf
A$ is full-rank over $\mbb C$, which is obviously easier.

\section{Proposed Cooperation Strategies}
\label{sec:prop-coop-stat}
In this section we propose two strategies for cooperation between the destination
and the relay. In the first strategy, the relay helps the destination by
providing its local ``best'' linear combination without taking into account
whether the resulting coefficient matrix is full rank or not. In the second
strategy, the relay assists the destination by forwarding  the ``best'' linear combination
that ensures the resulting coefficient matrix is full rank. Since the relay
needs to know the linear combinations that the destination possesses, a
sufficient amount of feedback is needed in the second strategy. 

Before the cooperation stage begins, the destination and the relay find
$M$ linearly independent coefficient vectors. Note that at this pre-cooperation stage,
the corresponding linear combinations are not yet decoded. Let $\mcal A_d =
\{\ad 1, \dots, \ad M\}$ and $\mcal A_r =
\{\ar 1, \dots, \ar M\}$ be the sets of coefficient vectors found by the
destination and the relay, respectively. The elements of $\mcal A_d$ and $\mcal
A_r$ are sorted based on the resulting computation rates. Let
$\mbf h_{sd} = [ h_{s_1d}, \dots,  h_{s_Md}]$  and $\mbf h_{sr} = [
h_{s_1r}, \dots,  h_{s_Mr}]$, and define
\begin{align}
  R_{\cp,d}^{(m)} &\triangleq R_{\cp} (\ad m, \mbf h_{sd}, g_{sd}),\label{eq:comp-rate-redefined-sd} \\
  R_{\cp,r}^{(m)} &\triangleq R_{\cp} (\ar m, \mbf h_{sr}, g_{sr}).\label{eq:comp-rate-redefined-sr}
\end{align}
The coefficient vectors $\ad 1,\dots, \ad M$ and $\ar 1, \dots, \ar M$ are
respectively sorted such that
$R_{\cp,d}^{(1)} \geq \cdots \geq R_{\cp,d}^{(M)}$ and $R_{\cp,r}^{(1)} \geq
\cdots \geq R_{\cp,r}^{(M)}$. Moreover, the elements of $\mcal A_d$ and $\mcal
A_r$ are integer vectors that provide $M$ successive minima corresponding to the matrix $\mbf B$ described in
\eqref{eq:marc-chol-decomp}. This implies that $\ad 1$ and $\ar 1$ are the local
optimal coefficient vectors at the destination and the relay, respectively.

Using these $\mcal A_d$ and $\mcal A_r$, we propose two cooperation strategies as follows.

\subsection{Limited Feedback Strategy}
\label{sec:limited-feedback}
The first strategy is simple, yet it outperforms the existing
cooperation strategies in the literature. In this strategy, there are two steps for
decoding the transmitted messages. The first is, upon receiving
$\mbf y_d^{(1)}$, the destination directly attempts to decode the transmitted
information messages without the help of the relay. We refer to this step of
decoding as \textit{direct decoding}. Specifically, using the integer
coefficient vectors in $\mcal A_d$, the destination decodes the
corresponding $M$ linear combinations. Let $\uhd 1, \dots, \uhd M$ be the
decoded linear combinations and $\mbf A_d = [\ad 1, \dots, \ad M]^T$. Based on
$\uhd 1, \dots, \uhd M$ and $\mbf A_d$, the destination attempts to decode the transmitted messages by
solving the resulting equation system similar to \eqref{eq:final-decoding}. 
If the destination successfully decodes the transmitted messages, the
sources can transmit the next messages. Otherwise, the destination broadcasts 
feedback to the sources and the relay, asking the sources to wait and the relay
to help the decoding. The feedback size is only $1$-bit and
is assumed to always be received correctly.

The second step of decoding, to which we refer as \textit{cooperative
  decoding}, is carried out when the relay receives feedback from 
the destination. The relay chooses its local best coefficient vector $\ar 1$,
decodes the corresponding linear combinations and forwards it to the destination.
Let $\uhr 1$ be the linear combination forwarded by the relay. The destination
now has an additional linear combination $\uhr 1$ with coefficient vector $\ar 1$. Let
$\mbf A_{\tr{cop}} = [\ar 1, \ad 1, \dots, \ad {M-1}]^T$. Based on $\uhr 1,
\uhd 1, \dots, \uhd {M-1}$ and $\mbf A_{\tr{cop}}$, the destination then again
decodes the transmitted messages. Note that $\mbf A_{\tr{cop}}$ may
not be full-rank which will prevent the destination from decoding the
information messages correctly. 

\subsection{Sufficient Feedback Strategy}
\label{sec:sufficient-feedback}

The second strategy is similar to the first in the sense that it also uses
two decoding steps. The first step is the same as the limited
feedback strategy. The destination attempts to \textit{directly} decode the information messages
using its own linear combinations. Using $\uhd 1, \dots, \uhd M$ and $\mbf A_d$ the
destination recovers $\w 1, \dots, \w M$ by solving an equation system
corresponding to \eqref{eq:final-decoding}. If direct decoding succeeds, the
sources can transmit their next information messages. Otherwise, the destination
sends feedback to the relay.  The feedback must contain
information about the $M-1$ best integer coefficient vectors of the destination,
i.e., $\ad 1, \dots, \ad {M-1}$. Besides perfectly received, it is also assumed that the feedback size is negligible compared to the
amount of information that can be transmitted within one coherence time. In practice, the feedback will require at least $M(M-1) \log p$ bits.

The relay selects a coefficient vector that ensures that the resulting coefficient
matrix is full rank while keeping the achievable rate as high as possible. Specifically, let $\ar *$ be the integer coefficient vector selected by the relay. Let $\mbf A_{rl} = [\ar l, \ad 1, \dots, \ad {M-1}]^T$. The relay must select $\ar *$
such that
\begin{align}
  \label{eq:suf-coef-vect}
  \ar * = \argmax_{\substack{\ar l \in {\ar 1, \dots, \ar M},\\ \rank{\mbf A_{rl}}=M}} R_{\cp,r}^{(l)}.
\end{align}
Subsequently, the relay decodes the linear combination of the messages
that corresponds to the selected coefficient vector $\ar *$ and then forwards it to
the destination. Now, because the destination has enough linear combinations, it
can re-decode the information messages by solving the resulting linear equation
system according to \eqref{eq:final-decoding}.

\section{Performance Analysis}
\label{sec:performance-analysis}

\subsection{Limited Feedback Strategy}
\label{sec:anal-lim-fb}
We start from the performance analysis of the limited feedback (lim-FB)
strategy. In this strategy, the destination has two possible ways of decoding
the transmitted messages, direct decoding and cooperative decoding.
In direct decoding, the destination attempts to decode the transmitted
messages by itself. Specifically, it computes linear combinations with
coefficients $\ad 1, \dots , \ad M$ and solves the corresponding linear equation
system. Let $e_1$ be the outage event for direct decoding. $e_1$ is defined as
\begin{align}
  e_1 &\triangleq \left\{ \min_{m \in \{1,\dots, M\}} R_{\cp,d}^{(m)} < R \right\} \label{eq:lim-e1-1} \\
      &= \left\{ R_{\cp, d}^{(M)} < R \right\}, \label{eq:lim-e1-2}
\end{align}
where $R$ is the coding rate employed by the sources and the relay. Note that
\eqref{eq:lim-e1-2} is due to the fact $R_{\cp,d}{(1)} \geq \cdots \geq R_{\cp,
  d}^{(M)}$, see Section~\ref{sec:prop-coop-stat}. Intuitively, we can think that the
outage event for direct decoding is determined by the worst coefficient vector $\ad M$.

In cooperative decoding, the relay forwards its local best linear
combination and the destination uses its $M-1$ best linear combinations and
solves the resulting linear equation system to decode the transmitted messages.
In order for cooperative decoding to succeed, all the linear combinations
have to be correctly decoded and the resulting coefficient matrix has to be full
rank. Let $e_2$ be the outage event for the cooperative decoding and $\mbf A$
be the resulting coefficient matrix. The outage event during cooperative decoding is defined
as 
\begin{align}
  e_2 &\triangleq \big\{ \min_{m \in \{1, \dots, M-1 \}}  R_{\cp, d}^{(m)} < R\big\}  \cup \big\{ R_{\cp, r}^{(1)} < R \big\} \nonumber \\
  & \quad \cup \big\{ \Rrd < R \big\} \cup \big \{ \rank{\mbf A} < M \big \}  \nonumber \\ 
      &=  \big\{ R_{\cp, d}^{(M-1)} < R\big\} \cup \big\{ R_{\cp, r}^{(1)} < R \big\} \cup \big\{ \Rrd < R \big\} \nonumber \\
  & \quad  \cup \big \{ \rank{\mbf A} < M \big \} \label{eq:lim-e1-e2-1} 
\end{align}
where $\Rrd = \log(1 + \abs{h_{rd}}^2 \gamma_{rd})$ is the achievable rate of
the point-to-point relay-destination link. 

Let $\Pdef \triangleq \Pr (\rank{\mbf A} < M)$. In the end, the destination 
fails to decode the transmitted messages if and only if both
direct and the cooperative decodings fail. Therefore, the outage probability
is given by
\begin{align}
  \Pout &\triangleq \Pr (e_1 \cap e_2)  \\
        &= \Pr \Big(\{ R_{\cp, d}^{(M)} < R \}  \cap  \big (\{ R_{\cp, d}^{(M-1)} < R\} \cup \{ R_{\cp, r}^{(1)} < R \} \nonumber \\
        & \quad \cup \{ \Rrd < R \} \cup \{ \rank{\mbf A} < M  \} \big ) \Big) \nonumber \\ 
        &\overset{(a)}{\leq} \Pr \big(\{ R_{\cp, d}^{(M-1)} < R \} \big ) +  \Pr \big (\{ R_{\cp, d}^{(M)} < R\} \nonumber \\
        & \quad  \cap \{ R_{\cp, r}^{(1)} < R \} \big ) + \Pr \big ( \{ R_{\cp, d}^{(M)} < R \} \cap \{ \Rrd < R \} \big ) \nonumber \\
        & \quad + \Pr \big ( \{ R_{\cp, d}^{(M)} < R \} \cap \{ \rank{\mbf A} < M  \} \big ) \big) \nonumber \\ 
        &\overset{(b)}{\leq} \Pr \big( R_{\cp, d}^{(M-1)} < R \big ) +  \Pr \big ( R_{\cp, d}^{(M)} < R \big) \Pr \big ( R_{\cp, r}^{(1)} < R \big ) \nonumber \\
        & \quad + \Pr \big ( R_{\cp, d}^{(M)} < R \big ) \Pr \big( \Rrd < R \big ) \nonumber \\
        & \quad + \min \big\{\Pr \big ( R_{\cp, d}^{(M)} < R \big ), \Pdef \big \} \label{eq:lim-fb-out}
\end{align}
where $(a)$ is due to the union bound and $(b)$ is because the channels of 
sources-destination, sources-relay, and relay-destination are independent. The
last part of $(b)$ is due to the Fréchet bound \cite{Frechet35}.

Besides outage probability, we are also interested in the diversity order achieved by the proposed strategies. Let us recall the definition of diversity order \cite{ZhengT03} achieved by a system. 
\begin{mydefinition}[Diversity order]
  For a system with outage probability $\Pout$, the diversity order of the
  system $d$ is defined as
  \begin{align}
    \label{eq:div-order}
    d \triangleq - \lim_{\gamma \rightarrow \infty} \frac {\log \Pout}{\log \gamma},
  \end{align}
  where $\gamma$ is the average SNR of the channels.
\end{mydefinition}
For simplicity, we also use an alternative form $\Pout \doteq \gamma^d$ to represent \eqref{eq:div-order}. The symbol $\doteq$ indicates the asymptotic equality for $\gamma \rightarrow \infty$.
The relation $\dotleq$ indicates a similar meaning.

Equation \eqref{eq:lim-fb-out} shows that the overall outage probability of the lim-FB
strategy depends on the outage probability of the sources-destination, sources-relay, and relay-destination links,
and the probability of rank deficient coefficient matrix during the cooperative
decoding. Because the relay-destination is a point-to-point link, it has diversity order one, see \cite{Tse05fundamentalsof} and \cite{LanemanTW04}. For
the sources-destination and sources-relay links, we have to evaluate the outage probability with respect to decoding linear combinations at the destination and the relay. In the following lemma, we show that
the destination and the relay achieve full diversity order for decoding their local optimal linear combinations.

\begin{mylemma}
\label{lemma:cf-diversity}
Consider a compute-and-forward scheme with $M$ transmitters, one receiver, and i.i.d. Rayleigh fading channels. The receiver wants to decode a linear combination of transmitted messages. Let $\gamma$ and $\Pout$ be the average SNR and outage probability, respectively. The diversity order achieves by the receivers with respect to recovering a linear equation using its optimal integer coefficient vector is $M$, i.e., $\Pout \dotleq \gamma^{-M}$.
\end{mylemma}

\begin{proof}
See Appendix~\ref{sec:proof-cf-diversity}.
\end{proof}

\begin{figure}[t]
  \centering
  \begin{tikzpicture}
    \begin{semilogyaxis}[mlineplot,
      width=\linewidth,
      height=\linewidth,      
      xlabel = {$\gamma_{sd}$ (dB)},
      ylabel =  {Outage Probability},
      legend pos = south west,
      xmin=5, xmax=40, ymin=1e-6, ymax=1,
      ]
      \legend{First Equation, Second Equation, Third Equation}
      \addplot[curve1] table [x=snrdb, y expr=\thisrow{comb1}/\thisrow{trialNum}] {data/multi-linear-comb-three.dat};
      \addplot[curve2] table [x=snrdb, y expr=\thisrow{comb2}/\thisrow{trialNum}] {data/multi-linear-comb-three.dat};
      \addplot[curve3] table [x=snrdb, y expr=\thisrow{comb3}/\thisrow{trialNum}] {data/multi-linear-comb-three.dat};
    \end{semilogyaxis}
  \end{tikzpicture}
  \caption{The outage probability of the $M=3$ best linear equations of a \cf
    system with three transmitters. The first best equation achieves third-order
  diversity, while the second achieves second-order diversity. The last
  equation achieves first-order diversity.}
  \label{fig:outage-cf-3}
\end{figure}
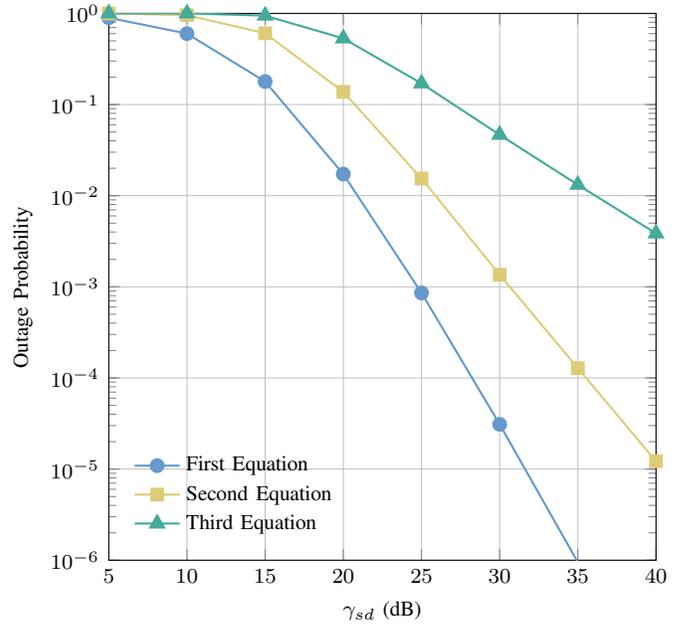
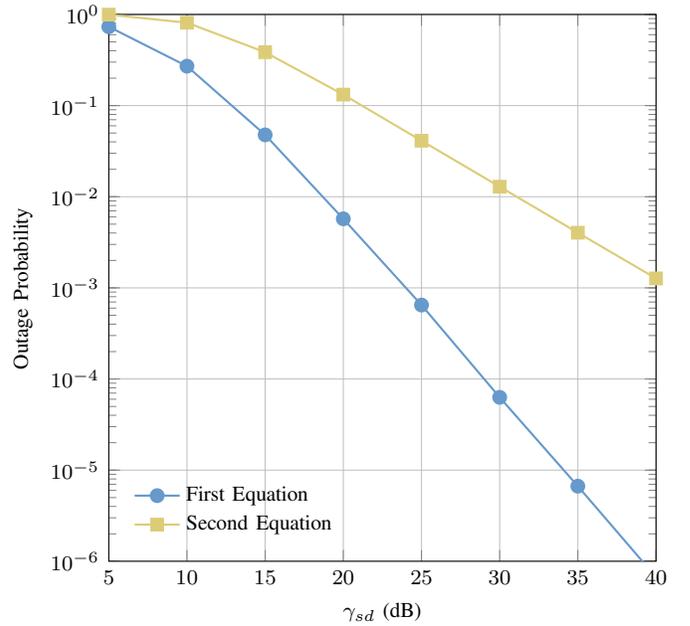
\begin{figure}
  \centering
  \begin{tikzpicture}
    \begin{semilogyaxis}[mlineplot,
      width=\linewidth,
      height=\linewidth,      
      xlabel = {$\gamma_{sd}$ (dB)},
      ylabel =  {Outage Probability},
      legend pos = south west,
      xmin=5, xmax=40, ymin=1e-6, ymax=1,
      ]
      \legend{First Equation, Second Equation}
      \addplot[curve1] table [x=snrdb, y expr=\thisrow{comb1}/\thisrow{trialNum}] {data/multi-linear-comb-two.dat};
      \addplot[curve2] table [x=snrdb, y expr=\thisrow{comb2}/\thisrow{trialNum}] {data/multi-linear-comb-two.dat};
    \end{semilogyaxis}
  \end{tikzpicture}
  \caption{The outage probability of the $M=2$ best linear equations of a \cf
    system with two transmitters. The first best equation achieves second-order
    diversity and the last equation achieves first-order diversity.}
  \label{fig:outage-cf-2}
\end{figure}
In addition to Lemma~\ref{lemma:cf-diversity} which shows the achieved diversity order of a \cf scheme with the first-best (local optimal) integer coefficient vector, we also need to know the achieved diversity order when the $(M-1)$-th and $M$-th best integer coefficients are selected. To this end, we provide numerical results evaluating the outage
probabilities of a \cf scheme with $M$ best linear coefficient vectors in Figs.~\ref{fig:outage-cf-3}~and~\ref{fig:outage-cf-2}. By
``best'' here, we mean the coefficient vectors that provide successive minima of
the resulting lattice $\mbf B$ in \eqref{eq:marc-chol-decomp}. Fig.~\ref{fig:outage-cf-3} shows outage probabilities
for each equation of a \cf system with three transmitters. Based on the slopes
of the outage probability curves, we can see that the first best linear equation
achieves diversity gain of order three which agrees with Lemma~\ref{lemma:cf-diversity}. The second best linear euqation achieves diversity
gain of order two and the last one achieves first-order
diversity gain. Similar results for the MARC with two transmitters are also shown in
Fig.~\ref{fig:outage-cf-2},
where the best linear equation achieves full
diversity order and the worst linear equation only achieves the first-order.
With these results, we conjecture that the $m$-th best linear equation of a \cf system
achieves diversity gain of order $M-m+1$.

\begin{figure}[t]
  \centering
  \begin{tikzpicture}
    \begin{semilogyaxis}[mlineplot,
      width=\linewidth,
      height=\linewidth,      
      xlabel = {$\gamma_{sd}$ (dB)},
      ylabel = {Rank Deficient Probability ($\Pdef$)},
      grid = major,
      legend pos = south west,      
      xmin = 5, xmax = 40, ymin = 5e-6, ymax=1e-1
      ]
      \legend{{$\delta_{sr} = .10$ ($+35.2$ dB)} ,{$\delta_{sr} = .25$ ($+21.193$ dB)}, {$\delta_{sr} = .50$ ($+10.596$ dB)}, {$\delta_{sr} = .75$ ($+4.3978$ dB)}}
      \addplot[curve1] table [x=sd_snrdb, y expr=\thisrow{rank_fail_num}/\thisrow{trial_num}] {data/rank-failure-prob/rank-fail-sr-delta-2src-01.dat};
      \addplot[curve2] table [x=sd_snrdb, y expr=\thisrow{rank_fail_num}/\thisrow{trial_num}] {data/rank-failure-prob/rank-fail-sr-delta-2src-025.dat};
      \addplot[curve3] table [x=sd_snrdb, y expr=\thisrow{rank_fail_num}/\thisrow{trial_num}] {data/rank-failure-prob/rank-fail-sr-delta-2src-05.dat};
      \addplot[curve4] table [x=sd_snrdb, y expr=\thisrow{rank_fail_num}/\thisrow{trial_num}] {data/rank-failure-prob/rank-fail-sr-delta-2src-075.dat};
    \end{semilogyaxis}
  \end{tikzpicture}
  \caption{Probability of rank deficient coefficient matrix ($\Pdef$) of MARC
    with two sources.}
  \label{fig:rank-fail-prob}
\end{figure}
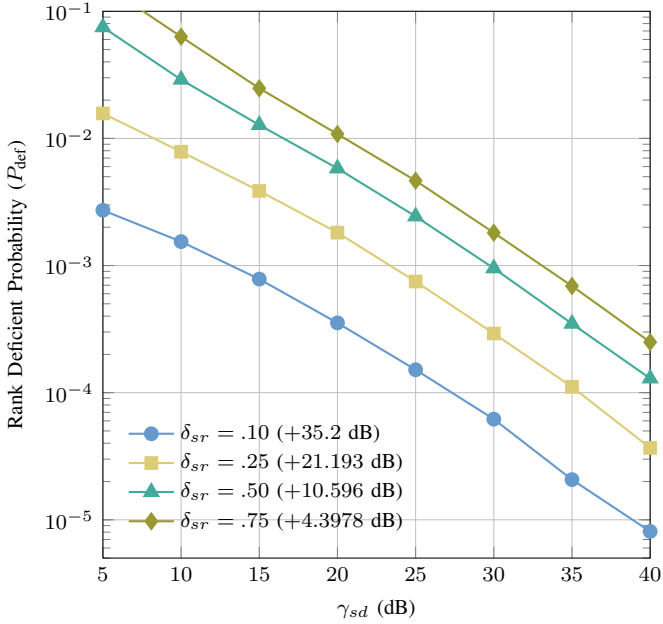

Now we are left with the probability of a rank deficient coefficient matrix $\Pdef$. To analyse the behavior of $\Pdef$, we present the rank deficient probability of the
coefficient matrices constructed during the cooperative decoding of the lim-FB
strategy in Fig.~\ref{fig:rank-fail-prob}. The numerical evaluation in Fig.~\ref{fig:rank-fail-prob} is performed by adjusting the position of the relay
relative to the sources $\delta_{sr}$ which impacts the average SNR. We assume
that the relay-destination link is perfect. It can be observed that the rank
deficient probability decreases as the position of the relay is closer to the
sources. Moreover, we can also observe from the slopes of the curves that the
rank deficient probability has an equivalent diversity gain of order less than
one. Let $k$ be the diversity order related to the rank deficient probability of the coefficient matrix. From the above results, we conjecture that $k<1$.

The outage probability of the lim-FB strategy now can be written as 
\begin{align}
  \Pout &\leq \Pr \big( R_{\cp, d}^{(M-1)} < R \big ) +  \Pr \big ( R_{\cp, d}^{(M)} < R \big) \Pr \big ( R_{\cp, r}^{(1)} < R \big ) \nonumber \\
        & \quad + \Pr \big ( R_{\cp, d}^{(M)} < R \big ) \Pr \big( \Rrd < R \big ) \nonumber \\
        & \quad + \min \big\{\Pr \big ( R_{\cp, d}^{(M)} < R \big ), \Pdef \big \} \nonumber \\
         &\overset{(a)}{\dotleq} \frac{\xi_1}{\gamma_{sd}^2} + \frac{\xi_2}{\gamma_{sd}} \frac{\xi_3}{\gamma_{sd}^M}+
           \frac{\xi_2}{\gamma_{sd}} \frac{\xi_4}{\gamma_{sd}} + \frac{\xi_2}{\gamma_{sd}}   \\
         &\dotleq \frac{\xi}{\gamma_{sd}}, \label{eq:lim-fb-diversity-order}
\end{align}
where $\xi, \xi_1, \xi_2, \xi_3 $, and $\xi_4$ are positive constants. In $(a)$,
because $k < 1$, in the high SNR regime, $\min \{\Pr ( R_{\cp, d}^{(M)} < R ),
\Pdef\} = \Pr (R_{\cp, d}^{(M)} < R)$. The above result indicates that
the lim-FB cannot achieve the full-diversity gain of the MARC. However, the
bound in \eqref{eq:lim-fb-diversity-order} is loose because of the Fréchet
bound. As we will see in the next section, the lim-FB nearly achieves diversity gain
of order two and its outage performance is significantly better compared to the
existing strategies.

\subsection{Sufficient Feedback Strategy}
\label{sec:anal-suf-fb}

The outage probability of the sufficient feedback (suf-FB) strategy is
similar to that of the lim-FB strategy. In the suf-FB strategy, there are also two
possible ways for the destination to decode the transmitted messages. The first
one is direct decoding, which is exactly the same as that of the lim-FB.
Therefore, the outage event for direct decoding in the suf-FB is also given by
\begin{align}
      e_1 &= \left\{ R_{\cp, d}^{(M)} < R \right\}. \label{eq:suf-e1-2}
\end{align}

The second one is cooperative decoding, where the relay select its best
linear combination $\ar *$ that is linearly independent of the first $M-1$ linear
combinations of the destination. Therefore, the resulting coefficient matrix is
always full rank. As a result, the outage event for cooperative decoding
depends only on the sources-destination, the sources-relay, and the
relay-destination links. Let $R_{\cp,r}^{(*)} \triangleq R_{\cp} (\ar *, \mbf
h_{sr}, g_{sr})$. The outage probability of the suf-FB strategy is defined as
\begin{align}
  e_2 &\triangleq \big\{ \min_{m \in \{1, \dots, M-1 \}}  R_{\cp, d}^{(m)} < R\big\} \cup \big\{ R_{\cp, r}^{(*)} < R \big\} \cup \big\{ \Rrd < R \big\} \nonumber \\ 
          &=  \big\{ R_{\cp, d}^{(M-1)} < R\big\} \cup \big\{ R_{\cp, r}^{(*)} < R \big\} \cup \big\{ \Rrd < R \big\} \label{eq:suf-e1-e2-1} 
\end{align}

Similar to the lim-FB strategy, the overall outage for the suf-FB strategy occurs if and only if both
direct and cooperative decodings fail. Therefore, the outage probability is
\begin{align}
  \Pout &\triangleq \Pr (e_1 \cap e_2)  \\
        &= \Pr \Big(\{ R_{\cp, d}^{(M)} < R \}  \cap  \big (\{ R_{\cp, d}^{(M-1)} < R\} \cup \{ R_{\cp, r}^{(*)} < R \}  \nonumber \\
        & \quad \cup \{ \Rrd < R \}  \Big) \nonumber \\ 
        &\overset{(a)}{\leq} \Pr \big( R_{\cp, d}^{(M-1)} < R \big ) +  \Pr \big ( R_{\cp, d}^{(M)} < R \big) \Pr \big ( R_{\cp, r}^{(*)} < R \big ) \nonumber \\
        & \quad + \Pr \big ( R_{\cp, d}^{(M)} < R \big ) \Pr \big( \Rrd < R \big ) \nonumber \\
        &\overset{(b)}{\leq} \Pr \big( R_{\cp, d}^{(M-1)} < R \big ) +  \Pr \big ( R_{\cp, d}^{(M)} < R \big) \Pr \big ( R_{\cp, r}^{(M)} < R \big ) \nonumber \\
        & \quad + \Pr \big ( R_{\cp, d}^{(M)} < R \big ) \Pr \big( \Rrd < R \big ) \label{eq:suf-fb-out}
\end{align}
where $(a)$ is due to union bound and in $(b)$ we bound $\Pout$ by selecting the worst linear combinations at the relay.

Using the results shown in the previous subsection, we can see that the
suf-FB strategy can achieve second-order diversity. Specifically, $\Pout$ can be written as
\begin{align}
  \Pout &\leq \Pr \big( R_{\cp, d}^{(M-1)} < R \big ) +  \Pr \big ( R_{\cp, d}^{(M)} < R \big) \Pr \big ( R_{\cp, r}^{(M)} < R \big ) \nonumber \\
         & \quad + \Pr \big ( R_{\cp, d}^{(M)} < R \big ) \Pr \big( \Rrd < R \big )\\
        &\dotleq \frac{\xi_1}{\gamma_{sd}^2} + \frac{\xi_2}{\gamma_{sd}}\frac{\xi_3}{\gamma_{sd}} + \frac{\xi_2}{\gamma_{sd}} \frac{\xi_4}{\gamma_{sd}} \\
         &\dotleq \frac{\xi}{\gamma_{sd}^{2}},
\end{align}
with other positive constants $\xi, \xi_1, \xi_2, \xi_3,$ and $ \xi_4$.

\section{Numerical Evaluation}
\label{sec:numerical-evaluation}

In this section, we provide results of computer simulations performed to evaluate the
performance of the proposed cooperation strategies, and compare them with approaches found
in the literature. Since we focus on the design of cooperation strategies for applying \cf to the MARC,
we mainly compare our proposed strategies to the approaches proposed by Soussi \textit{et al.}
\cite{SoussiZV14} and Insausti \textit{et al.} \cite{Insausti2019}. To the best
of our knowledge, \cite{SoussiZV14} and \cite{Insausti2019} are the only
works available in the literature that addressed the problem of applying \cf to the MARC.

Before going into the details, let us first briefly describe the approaches
proposed in \cite{SoussiZV14} and \cite{Insausti2019}. In \cite{SoussiZV14},
Soussi \textit{et al.} used a global optimization to choose linear
combinations at the relay and the destination. Given channel state
information (CSI) is known to all terminals, the relay and the destination
select their optimal linear combinations maximizing the achievable rate. The
relay then forwards its linear combination to the destination. Finally, having
sufficient linear combinations, the destination attempts to recover the
transmitted messages. It has been shown that this approach
achieves better achievable rate compared to other relaying strategies such as
\af and \df. However, it has a drawback that it requires greater communication
overhead due to the fact that all CSI is known to all nodes. Moreover, it can be proven that this approach does not achieve full-diversity gain as it has
a bottleneck in the link between the relay and the destination.  

In \cite{Insausti2019}, Insausti \textit{et al.} proposed an approach
where the relay is allowed to choose its best linear combination yielding
optimal computation rate and to forward it to the destination. The destination
then chooses linear combinations that are linearly independent of the one from
the relay and decodes the transmitted messages. If the decoding fails, the
destination computes one more linear combination from its received signal and
again decodes the transmitted messages. This approach is quite similar to our lim-FB strategy. Indeed, the two strategies achieve the same outage probability
performance as we will see later. They differ in the way they utilize the
transmission rounds. In the Insausti \textit{et al.~}approach, two transmission rounds are always
used. While in our proposed strategy, only one transmission round is used when possible to increase transmission efficiency. 

Now we describe conditions for the numerical evaluations. We assume all the sources have the same distance to the
destination and also to the relay. The distance from the sources to the
destination is denoted by $\delta_{sd}$, and to the relay is denoted by $\delta_{sr}$. The
relay has distance $\delta_{rd}$ to the destination. We
normalize $\delta_{sd} = 1$, and assume $\delta_{sr} + \delta_{rd} = \delta_{sd}$. See
the illustration in Fig.~\ref{fig:system-model}. The corresponding average SNRs
are calculated using a path-loss exponent equal to $3.52$
\cite{HolmaWCDMA,Youssef11}. In the simulations, we consider MARC with two
sources and transmission rate $R=2$. The performance is evaluated in three scenarios as follows.
\begin{enumerate}
\item \textit{First scenario:} The relay is closer to the sources than to the destination. Specifically, we
  set the distance from the sources to the relay $\delta_{sr} = 0.25$, while
  from the relay to the destination is $\delta_{rd} = 0.75$. This
  scenario is equivalent to the setting of average SNRs $\gamma_{sr} =
  \gamma_{sd} + 21.19$~dB and $\gamma_{rd} = \gamma_{sd} + 4.39$~dB. 
\item \textit{Second scenario:} The distance from
  the sources to the relay is equal to the distance from the relay to the
  destination, i.e., $\delta_{sr} = \delta_{rd} = 0.5$. In other words, the
  relay is half-way between the sources and the destination. With the same
  path-loss exponent, the resulting average SNRs 
  are  $\gamma_{sr} = \gamma_{rd} = \gamma_{sr} + 10.59$~dB.
\item \textit{Third scenario:} The relay is positioned closer to the destination than to the relay. In particular, we
  assume $\delta_{sr} = 0.25$ and $\delta_{rd} = 0.75$. As a result, the average $\gamma_{sr} =
  \gamma_{sd} + 4.39$~dB and $\gamma_{rd} = \gamma_{sd} + 21.19$~dB. 
\end{enumerate}

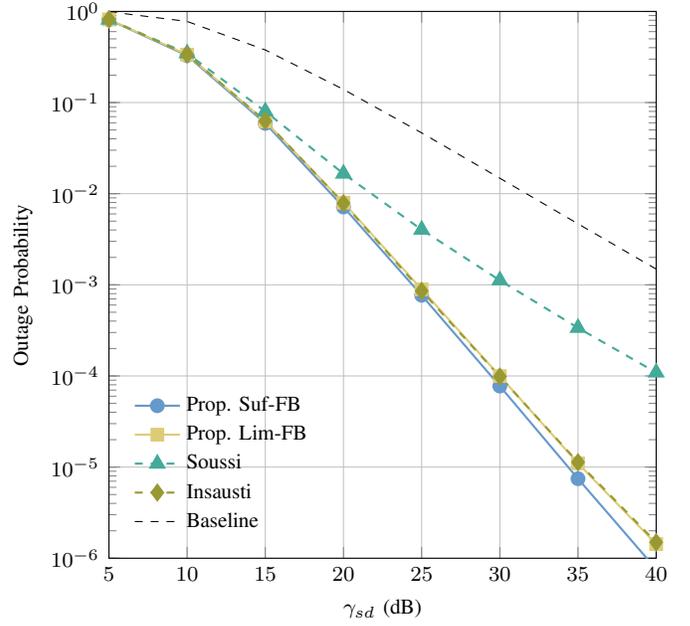
\begin{figure}[t]
  \centering
  \begin{tikzpicture}
    \begin{semilogyaxis}[mlineplot,
      width=\linewidth,
      height=\linewidth,      
      xlabel = {$\gamma_{sd}$ (dB)},
      ylabel =  {Outage Probability},
      legend pos = south west,
      xmin=5, xmax=40, ymin=1e-6, ymax=1,
      ]
      \legend{Prop. Suf-FB, Prop. Lim-FB, Soussi, Insausti, Baseline }
      \addplot[curve1] table [x=sd_snrdb, y expr=\thisrow{outage_num_snr}/\thisrow{trial_num}] {data/scenario-1/prop-two-cf-suf-fb-scen-01.dat};
      \addplot[curve2] table [x=sd_snrdb, y expr=\thisrow{outage_num_snr}/\thisrow{trial_num}] {data/scenario-1/prop-two-cf-lim-fb-scen-01.dat};
      \addplot[curve3, dashed] table [x=sd_snrdb, y expr=\thisrow{outage_num_snr}/\thisrow{trial_num}] {data/scenario-1/soussi-two-cf-scen-01.dat};
      \addplot[curve4, dashed] table [x=sd_snrdb, y expr=\thisrow{outage_num_snr}/\thisrow{trial_num}] {data/scenario-1/insausti-two-cf-scen-01.dat};
      \addplot[dashed] table [x=snrdb, y expr=\thisrow{outage_num_snr}/\thisrow{trial_num}] {data/two-user-marc-baseline.dat};
    \end{semilogyaxis}
  \end{tikzpicture}
  \caption{Outage probabilities of the two-source MARC in the first scenario, where the relay is closer to the sources.}
  \label{fig:outage-scen-1}
\end{figure}
\begin{figure}[t]
  \centering
  \begin{tikzpicture}
    \begin{semilogyaxis}[mlineplot,
      width=\linewidth,
      height=\linewidth,      
      xlabel = {$\gamma_{sd}$ (dB)},
      ylabel =  {Outage Probability},
      legend pos = south west,
      xmin=5, xmax=40, ymin=1e-6, ymax=1,
      ]
      \legend{Prop. Suf-FB, Prop. Lim-FB, Soussi, Insausti, Baseline }
      \addplot[curve1] table [x=sd_snrdb, y expr=\thisrow{outage_num_snr}/\thisrow{trial_num}] {data/scenario-2/prop-two-cf-suf-fb-scen-02.dat};
      \addplot[curve2] table [x=sd_snrdb, y expr=\thisrow{outage_num_snr}/\thisrow{trial_num}] {data/scenario-2/prop-two-cf-lim-fb-scen-02.dat};
      \addplot[curve3, dashed] table [x=sd_snrdb, y expr=\thisrow{outage_num_snr}/\thisrow{trial_num}] {data/scenario-2/soussi-two-cf-scen-02.dat};
      \addplot[curve4, dashed] table [x=sd_snrdb, y expr=\thisrow{outage_num_snr}/\thisrow{trial_num}] {data/scenario-2/insausti-two-cf-scen-02.dat};
      \addplot[dashed] table [x=snrdb, y expr=\thisrow{outage_num_snr}/\thisrow{trial_num}] {data/two-user-marc-baseline.dat};

    \end{semilogyaxis}
  \end{tikzpicture}
  \caption{Outage probabilities of the two-source MARC in the second scenario, where the relay is equidistant between the sources and destination.}
  \label{fig:outage-scen-2}
\end{figure}
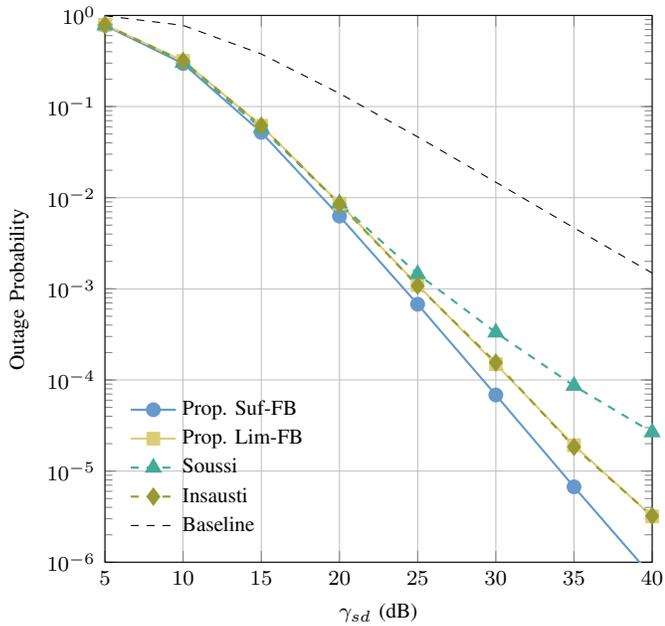
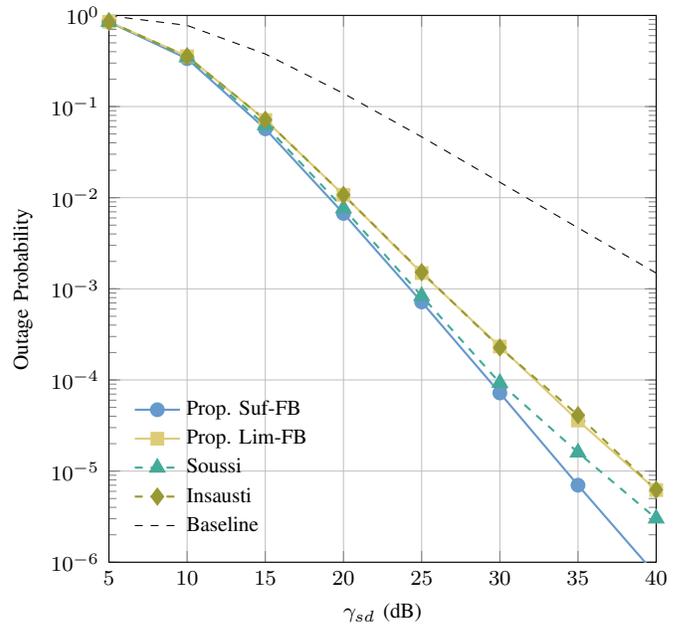
\begin{figure}[t]
  \centering
  \begin{tikzpicture}
    \begin{semilogyaxis}[mlineplot,
      width=\linewidth,
      height=\linewidth,      
      xlabel = {$\gamma_{sd}$ (dB)},
      ylabel =  {Outage Probability},
      legend pos = south west,
      xmin=5, xmax=40, ymin=1e-6, ymax=1,
      ]
      \legend{Prop. Suf-FB, Prop. Lim-FB, Soussi, Insausti, Baseline }
      \addplot[curve1] table [x=sd_snrdb, y expr=\thisrow{outage_num_snr}/\thisrow{trial_num}] {data/scenario-3/prop-two-cf-suf-fb-scen-03.dat};
      \addplot[curve2] table [x=sd_snrdb, y expr=\thisrow{outage_num_snr}/\thisrow{trial_num}] {data/scenario-3/prop-two-cf-lim-fb-scen-03.dat};
      \addplot[curve3, dashed] table [x=sd_snrdb, y expr=\thisrow{outage_num_snr}/\thisrow{trial_num}] {data/scenario-3/soussi-two-cf-scen-03.dat};
      \addplot[curve4, dashed] table [x=sd_snrdb, y expr=\thisrow{outage_num_snr}/\thisrow{trial_num}] {data/scenario-3/insausti-two-cf-scen-03.dat};
      \addplot[dashed] table [x=snrdb, y expr=\thisrow{outage_num_snr}/\thisrow{trial_num}] {data/two-user-marc-baseline.dat};

    \end{semilogyaxis}
  \end{tikzpicture}
  \caption{Outage probabilities of the two-source MARC in the third scenario, where the relay is closer to the destination.}
  \label{fig:outage-scen-3}
\end{figure}

The outage probability results for the first, second, and third scenarios are
presented in Figs.~\ref{fig:outage-scen-1}, \ref{fig:outage-scen-2}, and
\ref{fig:outage-scen-3}, respectively. Additionally, we present the
baseline outage probability for the case when the sources send their information to the
destination without a relay so that we can see how much improvement is gained
when the relay is employed. From here on we refer to the strategy proposed by
Soussi \textit{et al.} \cite{SoussiZV14} as \textit{Soussi strategy}, and
strategy proposed by Insausti \textit{et al.} \cite{Insausti2019} as
\textit{Insausti strategy}.

Figs.~\ref{fig:outage-scen-1} and \ref{fig:outage-scen-2} show that the Soussi strategy exhibits
the highest outage probability. This is because even though the linear combinations selected by
the relay and the destination are globally optimized, the destination can
correctly recover the transmitted message if and only if it correctly decodes its
own linear combination and the one from the relay. Therefore, if there is an
outage in either the source-relay link, the source-destination link, or the relay-destination
link, the final decoding at the destination will fail. This means that the relay
does not act as a helper, and rather, its presence is mandatory. Also, 
point-to-point communication from the relay to the destination can only achieve
first-order diversity gain, as can be confirmed in the numerical results, so the Soussi strategy suffers from a
bottleneck performance at the relay-destination link. This fact can be seen from the three
scenarios where the outage performance of the Soussi strategy gets better as the
distance of the relay to the destination gets smaller, i.e., $\gamma_{rd}$ gets
larger. 

In Figs.~\ref{fig:outage-scen-1} and \ref{fig:outage-scen-2}, it is shown that a
significant outage performance improvement over the Soussi strategy is achieved
by the Insausti strategy in the first and the second scenarios. For the third
scenario, even though at low SNR regime the Soussi strategy has lower outage probability, it
can be predicted that eventually the Insausti strategy is better in high SNR regime as
the slope of its outage probability curve is steeper. This improvement is a result of giving the
destination two possible ways of decoding the transmitting messages. The first
is with the help of the relay, and the second is by using linear combinations
decoded by itself. Thus, it can be thought that the relay acts as a helper
where its existence is not mandatory, i.e., it is possible for the destination to decode the
transmitted messages without the relay. It can also be observed that our
proposed lim-FB strategy achieves the same outage performance as the
Insausti strategy. This is because they are quite similar in the sense that the
destination has two possible ways for decoding the transmitted messages and treat
the relay as a useful helper. If we carefully observe the slopes of the outage
performance of the lim-FB and the Insausti strategies, they do not
achieve second-order diversity gain. The main reason behind this is that the local best linear
combination selected by the relay may not be linearly independent of the
$M-1$ best linearly combinations of the destination. We also observe that the
performance of the Insausti and the lim-FB strategies degrades
as the relay gets closer to the destination or as the average SNR from the sources to
the relay gets smaller. This is related to the probability of the rank
deficient coefficient matrix. As we have seen in Fig.~\ref{fig:rank-fail-prob},
the smaller the difference between $\gamma_{sd}$ and $\gamma_{sr}$, the higher
the probability of rank deficient coefficient matrix. Hence, for the lim-FB and
the Insausti strategies, it is better to place the relay closer to the sources. 

The best outage performance is achieved by the suf-FB
strategy. Based on the slopes of the curves shown in Figs.~\ref{fig:outage-scen-1}, \ref{fig:outage-scen-2}, and
\ref{fig:outage-scen-3}, one can see that the suf-FB strategy achieves the
second-order diversity gain. This agrees with our analysis in
Subsection~\ref{sec:anal-suf-fb}. The main reason for this is that the
destination has two possible ways in decoding the transmitted messages, the
direct and the cooperative decoding. Moreover, unlike in the lim-FB, the resulting coefficient
matrices in the suf-FB are guaranteed to always be full-rank. 

Next, we evaluate network throughput performance which is defined as the ratio
of the correctly received messages to number of transmission rounds utilized. For
the proposed strategies, because the second round of transmission is utilized
only when the direct decoding fails, the network throughput is defined as
\begin{align}
  \label{eq:prop-throughput}
  T_{\tr{prop}} = \frac{M (1 -\Pout )}{1 + \Pout^{\tr{dir}} },
\end{align}
where $\Pout^{\tr{dir}} \triangleq \Pr \{R_{\cp, d}^{(M)} < R\}$ is the outage probability
of the direct decoding. On the other hand, because the Soussi and the Insausti
strategies always use two transmission rounds, their network throughput is given
by
\begin{align}
  \label{eq:exist-throughput}
  T_{\tr{exist}} = \frac{M (1 -\Pout )}{2}.
\end{align}

\begin{figure}[t]
  \centering
  \begin{tikzpicture}
    \begin{axis}[mlineplot,
      width=\linewidth,
      height=\linewidth,      
      xlabel = {$\gamma_{sd}$ (dB)},
      ylabel =  {Throughput},
      legend pos = south east,
      xmin=5, xmax=40, ymin=0, ymax=2,
      ]
      \legend{Prop. Suf-FB, Prop. Lim-FB, Soussi, Insausti}
      \addplot[curve1] table [x=sd_snrdb, y expr=2*(1-\thisrow{outage_num_snr}/\thisrow{trial_num})/(1+\thisrow{direct_outage}/\thisrow{trial_num})] {data/scenario-2/prop-two-cf-suf-fb-scen-02.dat};
      \addplot[curve2] table [x=sd_snrdb, y expr=2*(1-\thisrow{outage_num_snr}/\thisrow{trial_num})/(1+\thisrow{direct_outage}/\thisrow{trial_num})] {data/scenario-2/prop-two-cf-lim-fb-scen-02.dat};
      \addplot[curve3, dashed] table [x=sd_snrdb, y expr=2*(1-\thisrow{outage_num_snr}/\thisrow{trial_num})/(2)] {data/scenario-2/soussi-two-cf-scen-02.dat};
      \addplot[curve4, dashed] table [x=sd_snrdb, y expr=2*(1-\thisrow{outage_num_snr}/\thisrow{trial_num})/(2)] {data/scenario-2/insausti-two-cf-scen-02.dat};
    \end{axis}
  \end{tikzpicture}
  \caption{Network throughput of the two-source MARC in the second scenario.}
  \label{fig:throughput-scen-2}
\end{figure}
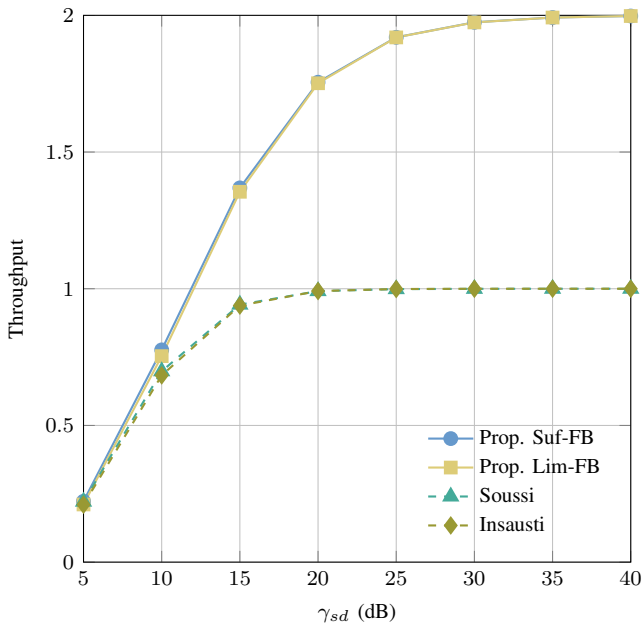
We found that in terms of network throughput, the performance of each strategy
in all scenarios is similar. Therefore, it is sufficient to only present the
network throughput performance
of one of the three scenarios; Fig.~\ref{fig:throughput-scen-2} presents the
network throughput of the second scenario. It is observed from Fig.~\ref{fig:throughput-scen-2}
that both the proposed strategies nearly achieve the maximum throughput of two
messages per transmission, while the existing strategies can only approach
maximum of one message per transmission. This is because the proposed strategies
requires less than two transmissions on average to deliver two messages. In
fact, in the high SNR regime, close to one transmission is required on average. On
the other hand, the existing strategies always utilize two transmission rounds
to deliver two messages. Therefore, the maximum network throughput they can
achieve is one message per transmission rounds. Thus, it can be concluded that the
proposed strategies have higher transmission efficiency compared to the existing
strategies.  

From the above results, it is clear that the suf-FB strategy achieves the best outage probability, diversity gain, and network throughput. These advantages indeed come with the cost a sufficient amount of overhead for the feedback. However, this
additional overhead is small enough compared to that of \cite{SoussiZV14}. Moreover, the feedback is only sent when
the destination fails to decode the transmitted messages by itself. Hence, in the
higher SNR regime, only a small amount of feedback is required. As an alternative, one may choose the lim-FB strategy that is better than the existing strategy in terms outage probability and network throughput with only one-bit feedback.

\section{Conclusions}
\label{3-sec:conclusion}
In this paper, we have studied the application of \cf to \marc (MARC). We proposed two
cooperation strategies between the relay and the destination. The proposed
strategies are opportunistic in the sense that they use transmission rounds as
few as possible to increase the transmission efficiency while improving outage
probability performance of the MARC. We have shown that both of the proposed
strategies improves network throughput remarkably, twice that of the existing
strategies \cite{SoussiZV14,Insausti2019}. It is also shown that both the proposed strategies always yields lower outage probability, independent of relay placement. Moreover, it is confirmed that the first strategy
called the lim-FB strategy achieves diversity gain close to the second-order,
which is a significant improvement over \cite{SoussiZV14}. A better outage
probability enhancement is achieved the second strategy, namely the suf-FB
strategy, where the full-diversity gain of the MARC is achieved. 

As future work, it is of interest to investigate the diversity multiplexing trade-off (DMT) of the MARC with \cf. Another direction is to allow the relay to help the destination several times, which can be regarded as an automatic repeat request (ARQ) scheme.
\appendices
\section{Proof of Lemma~\ref{lemma:cf-diversity}}
\label{sec:proof-cf-diversity}

Let $\mbf h = [h_1, \dots, h_M]$ be the channel coefficient vectors and $R$ be the coding rate employed by the transmitters. Without loss of generality, assume the geometric gain from the transmitters to the receiver is a unit, i.e., $g=1$ and the average SNR is $\gamma$. Let $\mbf a_{1}$ be the best (local optimal) integer coefficient vector selected by the receiver. The outage probability of decoding the linear combination corresponding to $\mbf a_1$ is defined as 
\begin{align}
    \Pout &\triangleq \Pr ( \Rcp(\mbf a_1, \mbf h, \mbf g) < R)\\
        &= \Pr \Bigg( \log^+ \bigg( \frac{1}{\mbf a_1^H \mbf M \mbf a} \bigg) < R \Bigg)\\
        &= \Pr \Bigg( \log^+ \bigg( \frac{1}{\norm{\mbf B \mbf a_1}^2}\bigg) < R \Bigg)\\
        &= \Pr \Bigg( \norm{\mbf B \mbf a_1}^2 > 2^{-R}\Bigg), \label{eq:appendix-1-bound-1}
\end{align}
where $\mbf M$ and $\mbf B$ are defined in \eqref{eq:marc-M-mat} and \eqref{eq:marc-chol-decomp}. 

As described in Section~\ref{sec:marc-comp-lin-comb}, $\mbf M$ is a positive definite matrix and $\mbf B$ is its Cholesky factorization. And because $\mbf a_1$ is the local optimal coefficient vector, $\norm{\mbf B \mbf a_1}$ is the first successive minimum or the minimum distance of the lattice generated by $\mbf B$. The minimum distance of a lattice is related to the determinant of its generator matrix by the Hermite's constant, see \cite{Nguyen10}. Let $\Psi_M$ be the Hermite's constant of dimension $M$. With $\Psi_M$, now we have a relation
\begin{align}
    \Psi_M \geq \frac{\norm{\mbf B\mbf a_1}^2}{\det{\mbf B}^{2/M}}.
\end{align} 
It can be shown that $\det{\mbf B} = 1 / ({1 + \gamma \norm{\mbf h}^2})^{M/2}$. Thus,
\begin{align}
    \frac{\Psi_M} {{1 + \gamma \norm{\mbf h}^2}} \geq \norm{\mbf B \mbf a}^2.
\end{align}
As a result, \eqref{eq:appendix-1-bound-1} can be written as
\begin{align}
    \Pout &= \Pr \Bigg( \norm{\mbf B \mbf a_1}^2 > 2^{-R}\Bigg) \\
    &\leq \Pr \Bigg( \frac{\Psi_M} {{1 + \gamma \norm{\mbf h}^2}} > 2^{-R}\Bigg) \\
    &\leq \Pr \Bigg(  \norm{\mbf h} < \frac{\Psi_M 2^{R} -1 }{\gamma} \Bigg )\\
    &\overset{(a)}{\leq}  \frac{\xi (\Psi_M2^{R} -1 )^M}{\gamma^M},
\end{align}
for a positive constant $\xi$, where $(a)$ is due to the generalization of \cite[Fact 2 in Appendix I]{LanemanTW04}. We then can show that the full-diversity order is achieved for decoding an optimal linear combination of a \cf scheme as
\begin{align}
     -\lim_{\gamma \rightarrow \infty} \frac{\log \Pout}{\log \gamma} \dotleq M,
\end{align}
or $\Pout \dotleq \gamma^{-M}.$
\bibliographystyle{IEEEtran}
\bibliography{HasanBibtex}

\end{document}